\documentclass[aps, superscriptaddress,nofootinbib, reprint]{revtex4-2}
\usepackage{graphicx}
\usepackage{xcolor}
\usepackage[%
colorlinks=true,
linkcolor=blue,
citecolor=blue,
]{hyperref}
\usepackage{booktabs}
\usepackage{rotating}
\usepackage[marginal]{footmisc}
\usepackage{graphicx}
\usepackage{dcolumn}
\usepackage{bm}

\usepackage{amsmath}
\allowdisplaybreaks[4] 

\usepackage[T1]{fontenc} 
\usepackage{tensor}
\usepackage{braket}  
\usepackage{bm}      
\usepackage{upgreek}
\usepackage{extarrows}

\newcommand\MaE{\mspace{2mu}\mathrm{e}\mspace{2mu}} 
\newcommand\MaPI{\mspace{2mu}\uppi\mspace{2mu}} 
\newcommand\MaD{\,\mathrm{d}} 
\newcommand\MaI{\mathrm{i}} 

\begin{document}
\title{Suppression of scalar perturbations due to a heavy axion}

\author{Kai-Ge Zhang}
\email{zhangkaige21@mails.ucas.ac.cn}
\affiliation{International Centre for Theoretical Physics Asia-Pacific,
University of Chinese Academy of Sciences, 100190 Beijing, China}
\affiliation{Taiji Laboratory for Gravitational Wave Universe, University of
Chinese Academy of Sciences, 100049 Beijing, China}

\author{Jian-Feng He}
\email{hejianfeng@itp.ac.cn}
\affiliation{Institute of
Theoretical Physics, Chinese Academy of Sciences (CAS), Beijing 100190, China}
\affiliation{School of Physical Sciences, University of Chinese Academy of
Sciences, No.19A Yuquan Road, Beijing 100049, China}

\author{Chengjie Fu}
\email{fucj@ahnu.edu.cn}
\affiliation{Department of Physics, Anhui Normal University, Wuhu, Anhui 241002, China}

\author{Zong-Kuan Guo}
\email{guozk@itp.ac.cn}
\affiliation{Institute of
Theoretical Physics, Chinese Academy of Sciences (CAS), Beijing 100190, China}
\affiliation{School of Physical Sciences, University of Chinese Academy of
Sciences, No.19A Yuquan Road, Beijing 100049, China}
\affiliation{School of Fundamental Physics and Mathematical Sciences,
Hangzhou Institute for Advanced Study, University of Chinese Academy of
Sciences, Hangzhou 310024, China}


\begin{abstract}
A fast-rolling axion can transfer its kinetic energy to gauge fields via the Chern-Simons coupling, leading to copious production of gauge quanta during inflation.
The amplified gauge fields act as a source for both scalar and tensor perturbations. 
In this work, we propose a mechanism for suppressing scalar perturbations while sourcing strong tensor perturbations.
We present an implementation of such a mechanism, demonstrating that sourced tensor perturbations are expected to be detected by upcoming next-generation CMB experiments.
\end{abstract}
\maketitle


\section{Introduction}
\label{sec: introduce}

Inflation is a natural extension of the standard hot big bang theory \cite{Guth:1980zm, Sato:1980yn, Linde:1981mu, Albrecht:1982wi, Starobinsky:1980te}. It introduces a stage of exponential expansion in the early Universe, which resolves both the horizon problem and the flatness problem. The standard slow-roll inflation predicts nearly scale-invariant primordial gravitational waves (GWs) on large scales, directly linked to the energy scale of inflation \cite{Starobinsky:1979ty, Lyth:1984yz, Lyth:1996im, Baumann:2006cd, Boubekeur:2012xn}.
Current constraints from cosmic microwave background (CMB) polarization experiments, especially the latest results from Planck \cite{Tristram:2020wbi} and BICEP/Keck \cite{BICEP2:2018kqh}, place stringent upper bounds on the tensor-to-scalar ratio ($r < 0.032$ at 95\% CL) \cite{Tristram:2021tvh}. Next-generation experiments are expected to achieve sensitivities of $r \sim 10^{-3}$ \cite{Hazumi:2019lys, CMB-S4:2016ple}.
On smaller scales, pulsar timing arrays (EPTA \cite{EPTA:2023fyk, EPTA:2023gyr, EPTA:2023xxk}, NANOGrav \cite{NANOGrav:2015aud, NANOGRAV:2018hou}, SKA \cite{Carilli:2004nx, Janssen:2014dka}), ground-based interferometers (LIGO \cite{Harry:2010zz}, Virgo \cite{VIRGO:2014yos}), and space-based interferometers (LISA \cite{LISA:2017pwj}, Taiji \cite{Ruan:2018tsw}) are set to probe the nHz and kHz frequency bands with unprecedented precision.

Aside from vacuum fluctuations, GWs can also be generated by additional excited fields~\cite{Cook:2011hg, Senatore:2011sp, Barnaby:2012xt, Carney:2012pk, Yu:2023ity, Ananda:2006af, Baumann:2007zm, Kohri:2018awv, Fu:2019vqc, Domenech:2019quo, Domenech:2020kqm, Pi:2020otn}. However, producing a significant amount of GWs generally requires a special mechanism in which the source temporarily counteracts the dilution caused by cosmic expansion~\cite{Cook:2011hg, Mirbabayi:2014jqa, Barnaby:2012xt}.
Among the few successful frameworks proposed, axion-gauge field dynamics during inflation has emerged as one of the most promising mechanisms~\cite{Namba:2015gja, Campeti:2022acx, Barnaby:2012xt, Ozsoy:2017blg, Garcia-Bellido:2016dkw}.
As shown in multiple studies~\cite{Barnaby:2010vf, Barnaby:2011vw, Meerburg:2012id, Sorbo:2011rz, Cook:2011hg, Barnaby:2011qe, Linde:2012bt, Dimopoulos:2012av, Urban:2013spa}, the transient rolling of the axion induces a tachyonic instability in one helicity mode of the gauge field, characterized by a dimensionless parameter $\xi$. This instability leads to the exponential amplification of gauge-field fluctuations, scaling as $\propto \MaE^{\pi \xi}$. The excited gauge fields subsequently act as a source of chiral GWs through quadratic contributions to the stress-energy tensor.
On the other hand, produced gauge quanta not only source GWs but also generate scalar perturbations with significant non-Gaussianity~\cite{Barnaby:2010vf, Barnaby:2011vw, Barnaby:2011qe,Linde:2012bt}. Consequently, constraints on scalar perturbations also limit particle production and the associated sourced GW signals~\cite{Ozsoy:2017blg}. Planck 2015~\cite{Planck:2015zfm} bounds on the non-Gaussianity parameter $f_{NL}$ impose an upper limit on $\xi$ of $\xi \lesssim 2.5$ on CMB scales. Such small values of $\xi$ make the sourced GWs negligible compared to vacuum fluctuations.


In order to suppress scalar perturbations while still generating detectable GW signals, some mechanisms are proposed~\cite{Ferreira:2014zia, Namba:2015gja, Garcia-Bellido:2016dkw,Kume:2025lvz}.
For example, if an axion fast-rolls for several e-foldings and then subsequently decays, it can produce observable GW signals on CMB or interferometer scales~\cite{Ferreira:2014zia, Namba:2015gja, Garcia-Bellido:2016dkw}.
More recently, Ref.~\cite{Kume:2025lvz} introduced an extension of axion-gauge field dynamics characterized by a non-canonical kinetic term. In the framework, a reduced sound speed of the axion field significantly suppresses scalar perturbations without diminishing sourced GWs. This creates a parameter space in which sourced GWs can dominate over the vacuum contribution, even within the context of axion inflation.

In this paper, we propose a mechanism to suppress scalar perturbations sourced by gauge quanta. The mechanism involves a heavy axion coupled to a light inflaton, with the axion also coupled to a $U(1)$ gauge field and having only a negligible influence on the inflaton’s dynamics. Because of the axion’s heavy mass, the field-space trajectory follows the partial minima of the potential. When the trajectory undergoes a rapid turn, the axion accelerates and triggers a tachyonic instability in the gauge field, leading to the exponential production of gauge quanta. These gauge quanta then source both scalar and tensor perturbations. For scalar perturbations associated with the axion, the heavy axion mass ensures that they are strongly suppressed. By contrast, tensor perturbations are unaffected by the axion mass and therefore remain unsuppressed.
We also provide a concrete implementation of our mechanism, in which the potential landscape is similar to that of two-field inflation models~\cite{Cespedes:2012hu, Pi:2017gih}. 
Our numerical results show that, 
the spectrum of sourced scalar perturbations is much smaller than the vacuum spectrum, while the sourced tensor spectrum exceeds its vacuum counterpart.

This paper is organized as follows. In Sec.~\ref{sec: mechanism}, we introduce the general idea of our mechanism, followed by the equations of motion (EoMs) for both the background quantities and the gauge field. We then present the formulas for computing the spectra of scalar and tensor perturbations. In Sec.~\ref{sec: implementation}, we provide a concrete implementation of our mechanism and compute the corresponding scalar and tensor spectra numerically. Finally, in Sec.~\ref{sec: conclusion}, we present our conclusions.


Throughout this paper, we adopt natural units with $\hbar = c = 1$ and define the reduced Planck mass as $M_{\mathrm{pl}}\equiv (8 \MaPI G)^{-1/2}$. The symbol $t$ denotes the cosmic time, while $\tau$ denotes the conformal time. A dot, as in $\dot{\phi} \equiv \mathrm{d} \phi / \mathrm{d} t$, indicates a derivative with respect to the cosmic time, and a prime, as in $\phi' \equiv \mathrm{d} \phi / \mathrm{d} \tau$ indicates a derivative with respect to the conformal time. The scale factor is $a(t)$ and the Hubble parameter is $H \equiv \dot{a} / a$. Our sign convention of metric is $(- + + +)$.


\section{Mechanism}
\label{sec: mechanism}

Our system is defined by the action
\begin{align}
  S=&\int \! \MaD^4x \, \sqrt{-g} \biggl[\
\frac{M_{\mathrm{pl}}^2}{2} R
- \frac{1}{2} \partial^\mu \phi \partial_\mu \phi-\frac{1}{2} \partial^\mu \chi \partial_\mu \chi
\nonumber \\
  & - V(\phi)-U(\phi,\chi) - \frac{1}{4} F^{\mu\nu} F_{\mu\nu}^{\phantom{\mu\nu}}
- \frac{\alpha}{4f} \chi F^{\mu\nu} \widetilde{\vphantom{\nu}F}_{\mu\nu}
\biggr],
\label{eq: L_axion}
\end{align}
where $\phi$ is the inflaton with potential $V(\phi)$, and is coupled to the axion field $\chi$ through $U(\phi,\chi)$. The evolution of the inflaton is mainly governed by $V(\phi)$, while the effect of the coupling $U(\phi,\chi)$ on the dynamics of $\phi$ is negligible. The tensor $F_{\mu\nu} \equiv \partial_{\mu}A_{\nu} - \partial_{\nu} A_{\mu}$ denotes the field strength of the U(1) gauge field, and its dual is defined as $\tilde{F}^{\mu\nu} \equiv \frac{1}{2}\eta^{\mu\nu\alpha\beta}F_{\alpha\beta}/\sqrt{-g}$, with the totally antisymmetric tensor $\eta^{\mu\nu\alpha\beta}$ normalized by $\eta^{0123} = 1$.
The parameter $f$ is the axion decay constant with mass dimension one, while $\alpha$ is a dimensionless parameter that controls the strength of the Chern-Simons interaction term $\frac{\alpha}{4 f} \chi \tilde{F}^{\mu\nu} F_{\mu\nu}$.

\begin{figure}[tbp]
  \includegraphics[width=.5\textwidth]{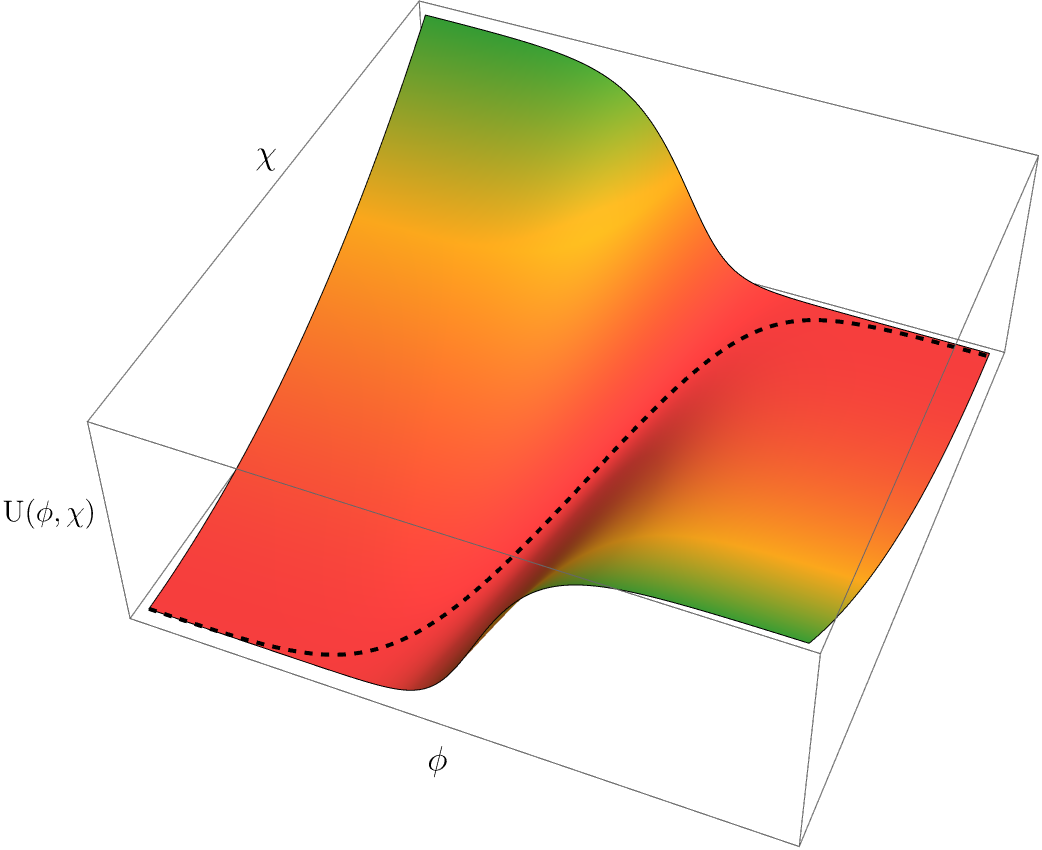}
  \caption{
  Schematic illustration of the coupling $U(\phi,\chi)$. The color denotes the value of $U(\phi,\chi)$, varying from red at low values to green at high values. The black dashed trajectory superimposed on the surface traces the partial-minimum valley, defined by the condition $\partial_{\chi} U(\phi,\chi) = 0$.
  }
  \label{fig: effective_potential}
\end{figure}

During the evolution, if the axion undergoes a fast-roll stage, gauge quanta are produced that source both scalar and tensor perturbations (see Sec.~\ref{sec: dynamics} for details). To suppress scalar perturbations while preserving the tensor perturbations, we consider a scenario in which the axion field is heavy. In this case, the inverse decay channel $A + A \to \delta \chi$ is suppressed, and gauge quanta no longer source scalar perturbations associated with $\delta \chi$.
However, during inflation, a heavy field typically settles at the minimum of its potential. Even if it is initially displaced, it quickly returns and oscillates around this point. Since gauge quanta production is proportional to the axion’s velocity, if the heavy axion evolves in isolation, its rapid deceleration prevents significant amplification of the gauge fields, thereby preventing the sourcing of large GWs.
To resolve this problem, we introduce a coupling $U(\phi,\chi)$ in the action~\eqref{eq: L_axion}, characterized by a curved valley structure, as illustrated in Fig.~\ref{fig: effective_potential}. This curved structure arises from the mixing between $\phi$ and $\chi$, i.e., $U(\phi,\chi) \neq U_1(\phi) + U_2(\chi)$. Without such mixing, the trajectory reduces to a straight line. Physically, this coupling can be interpreted as an effective potential that provides an effective mass $m_{\chi}$ for the axion, defined by
\begin{equation}
  \label{eq: m_chi}
  m_{\chi}^{2} \equiv U_{,\chi\chi}(\phi, \chi),
\end{equation}
where the comma denotes a partial derivative. We also define the similar quantity for $\phi$ as
\begin{equation}
  \label{eq: m_phi}
  m_{\phi}^{2} \equiv U_{,\phi\phi}(\phi, \chi).
\end{equation}


In this system, the axion $\chi$ acquires a heavy effective mass $m_{\chi} \gg H$ during its evolution and therefore remains on the trajectory of the partial minimum (i.e., $U_{,\chi}(\phi, \chi) = 0$).
On the other hand, since the influence of $U(\phi,\chi)$ on the inflaton is negligible, the inflaton still slowly rolls along its potential. The slow-roll condition implies that $m_{\phi}^{2} + V_{,\phi\phi} \ll H^{2}$, which in turn implies that $m_{\phi}$ is much smaller than the Hubble scale, $m_{\phi} \ll H$. Therefore, this system exhibits the hierarchy
\begin{equation}
m_{\chi} \gg H \gg m_{\phi}.
\end{equation}
As a result, $\chi$ is slowly displaced by the motion of $\phi$ but remains tightly bound to the partial-minimum trajectory due to its large mass. This trajectory is depicted by the black dashed line in Fig.~\ref{fig: effective_potential}.
In particular, when the angle between the direction of motion and the $\chi$-axis is small, the axion field enters a fast-roll phase, triggering a tachyonic instability and producing a large amount of gauge quanta. This stage corresponds to the middle segment of the black dashed line in Fig.~\ref{fig: effective_potential}.


\subsection{Dynamics}
\label{sec: dynamics}

From the action given in Eq.~\eqref{eq: L_axion}, along with the spatially flat Friedmann-Lemaître-Robertson-Walker metric, $\mathrm{d}s^{2} = -\mathrm{d}t^{2} + a^{2}(t) \mathrm{d}\bm{x}^{2}$, we can derive the following background equations,
\begin{align}
  \label{eq: a_eom_1}
  & H^{2} = \frac{1}{3 M_{\mathrm{pl}}} \rho, \\
  \label{eq: a_eom_2}
  & \frac{\ddot{a}}{a} + \frac{1}{2} \left( \frac{\dot{a}}{a}
  \right)^{2}
  = - \frac{1}{2 M_{\mathrm{pl}}} P, \\
  \label{eq: phi_eom}
  & \ddot{\phi} + 3 H \dot{\phi} + V_{,\phi} + U_{,\phi} = 0, \\
  \label{eq: chi_eom}
  & \ddot{\chi} + 3 H \dot{\chi} + U_{,\chi}
  = \frac{\alpha}{f} \braket{\bm{E} \cdot \bm{B}},
\end{align}
where the total energy density $\rho$ and the pressure $P$ are given by
\begin{align}
\label{eq: rho_tot}
  \rho =& \frac{1}{2} \dot{\phi}^{2}
  + \frac{1}{2} \dot{\chi}^{2}
  + V(\phi) + U(\phi,\chi) + \frac{1}{2}\braket{\bm{E}^{2} + \bm{B}^{2}} , \\
\label{eq: p_tot}
  P =& \frac{1}{2} \dot{\phi}^{2} + \frac{1}{2} \dot{\chi}^{2}
 - V(\phi) - U(\phi,\chi) + \frac{1}{6} \braket{\bm{E}^{2} + \bm{B}^{2}}.
\end{align}
Here, $\bm{E}$ and $\bm{B}$ are the electric and magnetic fields associated with the gauge field, defined by \cite{Barnaby:2010vf,Sorbo:2011rz,Garcia-Bellido:2023ser}
\begin{align}
  \label{eq: E_def}
  & E_{i}(t) = - \dot{A}_{i} / a, \\
  \label{eq: B_def}
  & B_{i}(t) = \epsilon_{ijk} \partial_{j} A_{k} / a^{2}.
\end{align}
The angle bracket $\langle \cdots \rangle$ in Eqs.~\eqref{eq: chi_eom}-\eqref{eq: p_tot} denotes the ensemble average of the fields, which encodes backreaction of the quantum field on the classical background~\cite{Ballardini:2019rqh, Peloso:2022ovc, Domcke:2020zez, Byrnes:2011aa, Cheng:2015oqa, Notari:2016npn, DallAgata:2019yrr, Gorbar:2021rlt, Durrer:2023rhc, Garcia-Bellido:2023ser, vonEckardstein:2023gwk, Iarygina:2023mtj, Galanti:2024jhw, vonEckardstein:2025oic, Anber:2009ua, Barnaby:2011qe, Sobol:2019xls, He:2024bno, He:2025ieo, Zhang:2025cyd, Caravano:2022epk, Caravano:2024xsb, Sharma:2024nfu, Figueroa:2024rkr, Lizarraga:2025aiw, Jamieson:2025ngu, Caravano:2021bfn, Caravano:2022epk, Figueroa:2023oxc}.

We decompose the gauge field $A_{i}(\tau, \bm{x})$ in the Fourier space as
\begin{align}
  A_{i}(\tau, \bm{x})
  = \int \frac{\mathrm{d}^{3}k}{(2\pi)^{{3}/{2}}}
  \sum_{\lambda=+,-} \varepsilon_{i}^{\lambda}(\hat{k}) A^{\lambda}(\tau, \bm{k})
  \MaE^{i \bm{k} \cdot \bm{x}},
\label{eq: Aq_Fourier}
\end{align}
where $\lambda = +, -$ denotes the polarization, $\hat{k} \equiv \bm{k}/k$ is a unit vector, $k \equiv |\bm{k}|$ is the norm, and $\epsilon_{i}^{\lambda}(\hat{k})$ is the circular polarization vector basis satisfying
\begin{align}
\label{eq: circular_1}
  & \bm{\varepsilon}^{\lambda}(\hat{{k}})
  = \left( \bm{\varepsilon}^{\lambda}(-\hat{{k}}) \right)^*, \\
  & \left( \bm{\varepsilon}^{\lambda}(\hat{{k}}) \right)^*
  \cdot \bm{\varepsilon}^{\lambda'}(\hat{{k}})
  = \delta_{\lambda \lambda'}, \\
  &\bm{{k}} \cdot \bm{\varepsilon}^{\lambda}(\hat{{k}}) = 0, \\
\label{eq: circular_4}
  &\bm{{k}} \times \bm{\varepsilon}^{\lambda}(\hat{{k}})
  = (-\lambda) \mathrm{i} k \, \bm{\varepsilon}^{\lambda}(\hat{{k}}).
\end{align}
We perform canonical quantization by promoting the classical field $A^{\lambda}(\tau,\bm{k})$ to the operator $\hat{A}^{\lambda}(\tau,\bm{k})$,
\begin{equation}
\label{eq: Aq_quantization}
\hat{A}^{\lambda}(\tau,\bm{k}) = A^{\lambda}(\tau,\bm{k})\hat{a}_{\lambda}(\bm{k}) + \left( A^{\lambda}(\tau,-\bm{k}) \right)^{\!*} \hat{a}_{\lambda}^{+}(-\bm{k}),
\end{equation}
where $\hat{a}_{\lambda}(\bm{k})$ and $\hat{a}_{\lambda}^{+}(\bm{k})$ are annihilation and creation operators, respectively, satisfying the standard commutation relations,
\begin{align}
  \label{eq: a_commut_1}
  \left[ \hat{a}_{\lambda}(\bm{k}), \hat{a}_{\lambda'}^{\dagger}(\bm{k}') \right]
  &= \delta_{\lambda\lambda'} \delta^{(3)} (\bm{k} - \bm{k}'), \\
  \label{eq: a_commut_2}
  \left[ \hat{a}_{\lambda}^{\dagger}(\bm{k}), \hat{a}_{\lambda'}^{\dagger}(\bm{k}') \right]
  &= 0, \\
  \label{eq: a_commut_3}
  \left[ \hat{a}_{\lambda}(\bm{k}), \hat{a}_{\lambda'}(\bm{k}') \right]
  &= 0.
\end{align}
Then, by the definition of $\bm{E}$ and $\bm{B}$, one can obtain
\begin{align}
  \label{eq: ensemble_EB}
  & \braket{\bm{E} \cdot \bm{B}} =
  - \frac{1}{4 \MaPI^{2} a^{4}}
  \sum_{\lambda=\pm} \lambda \int_{0}^{\infty}
  \mathrm{d}k k^3
  \frac{\mathrm{d} }{\mathrm{d} \tau} \left|A^{\lambda}(\tau,\bm{k})\right|^{2}, \\
  \label{eq: ensemble_EE}
  & \langle E^{2}\rangle=\frac{1}{2\pi^{2}a^{4}}\sum_{\lambda=\pm}\int_{0}^{\infty}\mathrm{d}k\, k^{2} \left| \frac{\mathrm{d}A^{\lambda}(\tau,\bm{k})}{\mathrm{d}\tau} \right|^{2}, \\
  \label{eq: ensemble_BB}
  & \braket{B^{2}} = \frac{1}{2 \MaPI^{2} a^{4}}
  \sum_{\lambda=\pm} \int_{0}^{\infty} \mathrm{d}k k^{4}
  |A^{\lambda}(\tau,\bm{k})|^{2}.
\end{align}

Substituting the decomposition in Eqs.~\eqref{eq: Aq_Fourier} and~\eqref{eq: Aq_quantization} into the action in Eq.~\eqref{eq: L_axion}, and subsequently varying the action, yields the EoM for the mode functions of the gauge field~\cite{Sorbo:2011rz,Garcia-Bellido:2023ser},
\begin{equation}
\label{eq: Aq}
  \ddot{A}^{\pm}(\tau, \bm{k}) + H \dot{A}^{\pm}(\tau, \bm{k})
  + \left( \frac{k^{2}}{a^{2}} \mp \frac{\alpha}{f} \frac{k}{a} \dot{\chi} \right)
  A^{\pm}(\tau, \bm{k}) = 0,
\end{equation}
where the term $\frac{\alpha}{f} \frac{k}{a} \dot{\chi}$ comes from the Chern-Simons interaction $\chi F^{\mu\nu} \tilde{F}_{\mu\nu}$.
When the effective mass $( \frac{k^{2}}{a^{2}} - \lambda \frac{\alpha}{f} \frac{k}{a} \dot{\chi}(\tau) )$ becomes negative, the corresponding helicity mode $A^{\lambda}(\tau,\bm{k})$ undergoes tachyonic instability and grows exponentially, while the opposite helicity $A^{-\lambda}(\tau,\bm{k})$ remains at its vacuum value~\cite{Barnaby:2011vw,Ozsoy:2020ccy}. Both modes freeze once they are well outside the horizon. This parity-violating property imprints itself on the sourced GW energy spectrum, leading to the production of chiral GWs~\cite{Gluscevic:2010vv,Seto:2007tn,Smith:2016jqs, Domcke:2019zls}.

Without loss of generality, we focus on the case $\mathrm{sign}(\chi'(\tau)) = -1$, in which the enhanced mode is $A^{-}(\tau,\bm{k})$. It is also common to define a parameter $\xi$ by
\begin{equation}
  \xi \equiv - \frac{\alpha \dot{\chi}}{2 f H},
\end{equation}
which encodes the velocity of the axion field. The initial condition of mode function $A^{\lambda}(\tau,\bm{k})$ is determined by the Bunch-Davies vacuum,
\begin{equation}
   \left. A^{\pm}(\tau,\bm{k}) \right|_{- k \tau \gg 1}
  = \frac{1}{\sqrt{2k}} \MaE^{ - \MaI k \tau }.
\end{equation}

\subsection{Scalar perturbations}
\label{sec: scalar}

Expanding the scalar fields as $\phi(t, \bm{x}) = \phi_0(t) + \delta \phi(t, \bm{x})$ and $\chi(t, \bm{x}) = \chi_0(t) + \delta \chi(t, \bm{x})$, the gauge-invariant comoving curvature perturbation $\mathcal{R}$ can be expressed in terms of the field variables as~\cite{Ozsoy:2020ccy}
\begin{align}
\notag
  \mathcal{R}(\tau,\bm{k})
  & = -\frac{H}{\dot{\phi}_{0}^{2} + \dot{\chi}_{0}^{2}} \Bigl(
  \dot{\phi}_{0}\, \delta\phi^{\mathrm{F}}(\tau,\bm{k})
  + \dot{\chi}_{0}\, \delta\chi^{\mathrm{F}}(\tau,\bm{k}) \\
\label{eq: comoving_curvature_kspace}
  & + \dfrac{a}{M^{3}_{\mathrm{pl}}} \frac{i k_{i}}{k^{2}}
  \int \frac{\mathrm{d}^{3}\bm{q}}{(2\pi)^{3/2}}
  \epsilon_{ijk} E_{j}(\tau, \bm{k} - \bm{q}) B_{k}(\tau, \bm{q})
  \Bigr),
\end{align}
where the superscript $\mathrm{F}$ denotes that perturbations are defined in the flat gauge, and, for simplicity, we omit this notation hereafter.
By expanding the total action~\eqref{eq: L_axion} to third order, we obtain the following perturbation equations in momentum space, expressed in terms of the rescaled canonical variables $Q_{\phi} \equiv a\delta\phi$ and $Q_{\chi} \equiv a\delta\chi$~\cite{Barnaby:2012xt, Ozsoy:2017blg, Barnaby:2011vw},
\begin{align}
\label{eq: Qphi_perturbation}
\notag
  & \ddot{Q}_{\phi} + H \dot{Q}_{\phi} + \left( \frac{k^{2}}{a^{2}}
  - \frac{\ddot{a}}{a} - H^{2} + m_{\phi}^{2} \right) Q_{\phi} \\
  &= J_{\mathrm{D}} Q_{\chi} - \frac{a}{M^{2}_{\mathrm{pl}}}
  \frac{\dot{\phi}_{0}}{4 H}
  \int \frac{\mathrm{d}^{3}\bm{p}}{(2\pi)^{3/2}}
\notag
  \left(-1 + \frac{(p - |\bm{k} - \bm{p}|)^{2}}{k^{2}}\right) \\
  &\times \left(
    {E}_{i}(\tau,\bm{p}){E}_{i}(\tau,\bm{k}-\bm{p}) +
    {B}_{i}(\tau,\bm{p}){B}_{i}(\tau,\bm{k}-\bm{p})
  \right), \\
\label{eq: Qchi_perturbation}
\notag
  & \ddot{Q}_{\chi} + H \dot{Q}_{\chi} + \left( \frac{k^{2}}{a^{2}}
  - \frac{\ddot{a}}{a} - H^{2} + m_{\chi}^{2} \right) Q_{\chi} \\
  &= J_{\mathrm{D}} Q_{\phi} + a\dfrac{\alpha}{f}
  \int \dfrac{\mathrm{d}^{3} \bm{p}}{(2\pi)^{3/2}}
  E_{i}(\tau,\bm{k}-\bm{p}) B_{i}(\tau,\bm{p}).
\end{align}
Here, we omit the conventional slow-roll suppressed correction coefficients associated with the mass terms, while all metric perturbations are placed by matter field perturbations through the Einstein's equation, and introduce the gravitational mixing term $J_{\mathrm{D}}$,
\begin{align}
\notag
  J_{\mathrm{D}} \equiv &
  H^2 \biggl( -\frac{1}{H^{2}} \partial_{\chi} \partial_{\phi} U
  + \frac{\partial_{\phi} U}{H^2 M_{\mathrm{pl}}}
  \sqrt{2\epsilon_{\chi}} \\
\notag
  & - \frac{\alpha}{f} \frac{1}{\sqrt{2}}
  \frac{\braket{\bm{E} \cdot \bm{B}}}{H^2M_{\mathrm{pl}}}
  \sqrt{\epsilon_{\phi}}
  + \frac{\braket{\bm{E}^{2}+\bm{B}^{2}}}{3 H^2 M^{2}_{\mathrm{pl}}}
  \sqrt{\epsilon_{\phi} \epsilon_{\chi}} \\
\label{eq: effective_J}
  & -2 (3 - \epsilon_{\phi} - \epsilon_{\chi})
  \sqrt{\epsilon_{\phi}\epsilon_{\chi}} \biggr),
\end{align}
where $\epsilon_{\phi} \equiv \dot{\phi}^{2} / (2 M_{\mathrm{pl}}^{2} H^{2})$ and $\epsilon_{\chi} \equiv \dot{\chi}^{2} / (2 M_{\mathrm{pl}}^{2} H^{2})$ are the slow-roll parameters of the two fields, satisfying $\epsilon_{\phi}, , \epsilon_{\chi} \ll 1$ during the slow-roll stage.
It is now clear how the three different scalar perturbation components of $\mathcal{R}(\tau,\bm{k})$ evolve and convert during inflation.
First, the background axion field transfers part of its kinetic energy to the gauge fields through the Chern-Simons interaction, i.e., $\chi_{0} \to A$.
The amplified gauge field then sources the $\delta \chi$ sector via the inverse decay process $A +  A \to \delta \chi$, which corresponds to the second term on the RHS of Eq.~\eqref{eq: Qchi_perturbation}.
Meanwhile, the excited gauge field also sources $\delta \phi$ through the gravitational channel $A + A \to \delta \phi$, which is represented by the second term on the RHS of Eq.~\eqref{eq: Qphi_perturbation}.
However, since the axion mass is heavy, $m_{\chi}^{2} \gg H^{2}$, the term $m^{2}_{\chi} Q_{\chi}$ dominates the LHS of Eq.~\eqref{eq: Qchi_perturbation}. Thus, in the large mass limit, the derivative term on the LHS of Eq.~\eqref{eq: Qchi_perturbation} can be neglected, giving the approximation $m_{\chi}^{2} Q_{\chi} \sim J_{\mathrm{D}} Q_{\phi}$. Substituting this relation into the first term on the RHS of Eq.~\eqref{eq: Qphi_perturbation} yields $J_{\mathrm{D}} Q_{\chi} \sim (J_{\mathrm{D}}^{2} / m^{2}_{\chi}) Q_{\phi}$. Since $J_{\mathrm{D}} \sim H^{2}$ and $m^{2}_{\chi} \gg H^{2}$, it follows that $J_{\mathrm{D}} Q_{\chi} \ll H^{2} Q_{\phi}$. Therefore, the term $J_{\mathrm{D}} Q_{\chi}$ on the RHS of Eq.~\eqref{eq: Qphi_perturbation} is negligible compared to $H^{2} Q_{\phi}$ on the LHS. Furthermore, because the inflaton is light, $m^{2}_{\phi} \ll H^{2}$, the term $m^{2}_{\phi}$ on the LHS can also be neglected.
In conclusion, Eq.~\eqref{eq: Qphi_perturbation} for $Q_{\phi}$ can be reduced to
\begin{align}
\label{eq: X_Qphi_perturbation}
\notag
  & \ddot{Q}_{\phi} + H \dot{Q}_{\phi} + \left( \frac{k^{2}}{a^{2}}
  - \frac{\ddot{a}}{a} - H^{2} \right) Q_{\phi} \\
  &= \frac{a}{M^{2}_{\mathrm{pl}}}
  \frac{\dot{\phi}_{0}}{4 H}
  \int \frac{\mathrm{d}^{3}\bm{p}}{(2\pi)^{3/2}}
\notag
  \left(-1 + \frac{(p - |\bm{k} - \bm{p}|)^{2}}{k^{2}}\right) \\
  &\times \left(
    {E}_{i}(\tau,\bm{p}){E}_{i}(\tau,\bm{k}-\bm{p}) +
    {B}_{i}(\tau,\bm{p}){B}_{i}(\tau,\bm{k}-\bm{p})
  \right),
\end{align}
and because of the heavy axion mass, Eq. \eqref{eq: Qchi_perturbation} implies \cite{Cespedes:2012hu, Achucarro:2012sm, Achucarro:2012yr}
\begin{equation}
\label{eq: X_Qchi_perturbation}
\delta \chi(\tau, \bm{k})|_{-k\tau \ll 1} = 0.
\end{equation}

Since the second term in Eq.~\eqref{eq: comoving_curvature_kspace} scales as $\bm{E} \times \bm{B} \propto \rho_{\mathrm{EM}} \propto a^{-4}$, the contribution of this term is negligible at the end of inflation. Combined with Eq.~\eqref{eq: Qchi_perturbation}, curvature perturbations in Eq.~\eqref{eq: comoving_curvature_kspace} reduce to
\begin{equation}
\label{eq: R_end}
  \mathcal{R}(\tau,\bm{k})
  \approx -\frac{H}{\dot{\phi}_{0}} \delta\phi(\tau, \bm{k}).
\end{equation}
Using Eq.~\eqref{eq: X_Qphi_perturbation} along with Eq. \eqref{eq: R_end}, one can obtain the total scalar power spectrum,
\begin{align}
\notag
  & \mathcal{P}_{\mathcal{R}}(\tau, k)
  = \frac{H^4}{4 \MaPI^{2} \dot{\phi}_{0}^{2} M^{4}_{\mathrm{pl}}}
  + \frac{k^{3} H^2}{64 \MaPI^{4} \dot{\phi}_{0}^{2} M^{4}_{\mathrm{pl}}}
  \int_{0}^{\infty} q^{2} \mathrm{d} q \\
\notag
  & \times \int_{-1}^{1} \mathrm{d} u
  \Biggr[
  \left| \epsilon^{-}_{i}(\frac{\bm{k}-\bm{q}}{|\bm{k}-\bm{q}|}) \epsilon^{-}_{i}(\hat{q}) \right|^{2}
  \left(-1 + \left(\frac{q-|\bm{k}-\bm{q}|}{k} \right)^{2} \right)^{2} \\
\notag
  & \times \biggl| \int_{-\infty}^{0} \mathrm{d} \tau'
     \frac{\dot{\phi}}{H} \frac{G_{k}(\tau, \tau')}{a(\tau)a(\tau')}
  \biggl(
    A^{'-}(\tau', \bm{q})A^{'-}(\tau', \bm{k}-\bm{q})\\
\label{eq: PR}
    & + q |\bm{k} - \bm{q}|
    A^{-}(\tau', \bm{q})A^{-}(\tau', \bm{k} - \bm{q})
  \biggr) \biggr|^{2} \Biggr],
\end{align}
where the Green's function is given by
\begin{align}
  \notag
  G_{k}(\tau, \tau') &= \dfrac{1}{k^{3} \tau\tau'}
  \bigl[
    (1 + k^{2} \tau \tau') \sin(k (\tau - \tau')) \\
  \label{eq: green_func}
    & - k (\tau - \tau') \cos(k (\tau - \tau'))
  \bigr] \Theta(\tau - \tau').
\end{align}

\subsection{Tensor perturbations}
\label{sec: tensor_perturbation}

The metric with tensor perturbations is written as
\begin{equation}
\label{eq: ds^2}
\mathrm{d}s^{2} = a^{2}(\tau) \left[
    -\mathrm{d}\tau^{2} + \left( \delta_{ij} + h_{ij}(\tau, \bm{x}) \right) \mathrm{d}x^{i} \mathrm{d}x^{j}
\right],
\end{equation}
where $h_{ij}$ represents the transverse and traceless metric perturbations, i.e., $h_{ii} = \partial_{j}h_{ij} = 0$.
The EoM of $h_{ij}(\tau, \bm{k})$ is given by
\begin{equation}
\label{eq: h_perturbation}
  \ddot{h}_{ij} + 3H\dot{h}_{ij} - \frac{1}{a^{2}} \nabla^{2} h_{ij}
  = -\frac{2}{M^{2}_{\mathrm{pl}}}\hat{\Pi}_{ij}^{lm}(E_{l}E_{m}+B_{l}B_{m}),
\end{equation}
where the projector operator is defined as $\hat{\Pi}_{ij}^{lm} \equiv \Pi_{l}^{i} \, \Pi_{m}^{j} - \frac{1}{2} \, \Pi_{ij} \Pi^{lm}$ with $\Pi_{ij} \equiv \delta_{ij} - \partial_{i} \partial_{j} / \nabla^{2}$.
We then decompose tensor perturbations $\hat{h}_{ij}(\tau, \bm{x})$ in Fourier space as
\begin{equation}
\label{eq: hq_Fourier}
\begin{aligned}
& \hat{h}_{ij}(\tau, \bm{x}) \\
&= \frac{2}{M_{\mathrm{pl}}}\frac{1}{a(\tau)}
  \int \frac{\mathrm{d}^3 \bm{k}}{(2\pi)^{{3}/{2}}}
  \sum_{\lambda=+,-}
  \varepsilon_{i}^{\lambda}(\hat{k}) \varepsilon_{j}^{\lambda}(\hat{k})
  \hat{h}_{\lambda}(\tau, \bm{k})
  e^{i \bm{k} \cdot \bm{x}} \\
&= \frac{2}{M_{\mathrm{pl}}}\frac{1}{a(\tau)}
  \int \frac{\mathrm{d}^3 \bm{k}}{(2\pi)^{{3}/{2}}}
  \sum_{\lambda=+,-}
  \varepsilon_{i}^{\lambda}(\hat{k}) \varepsilon_{j}^{\lambda}(\hat{k}) \\
&\times \bigg( h_{\lambda}^{\mathrm{v}}(\tau, \bm{k}) \hat{b}(\bm{k})
  + \Big( h_{\lambda}^{\mathrm{v}}(\tau, -\bm{k}) \Big)^{\!*} \hat{b}^{\dagger}(-\bm{k})
  + \hat{h}^{\mathrm{s}}(\tau, \bm{k}) \bigg) e^{i \bm{k} \cdot \bm{x}},
\end{aligned}
\end{equation}
where $\hat{b}_{\lambda}(\bm{k})$ and $\hat{b}_{\lambda}^{\dag}(\bm{k})$ are the annihilation and creation operators, respectively, satisfying the standard commutation relations for $\hat{a}_{\lambda}(\bm{k})$ in Eqs.~\eqref{eq: a_commut_1}-\eqref{eq: a_commut_3}; $\epsilon_{i}^{\lambda}(\hat{k})$ is the circular polarization vector basis defined in Eqs.~\eqref{eq: circular_1}-\eqref{eq: circular_4}; and ${h}_{\lambda}^{\mathrm{v}}{}(\tau, \bm{k})$ is a homogeneous solution corresponding to the vacuum GWs, which satisfies
\begin{equation}
  h_{\lambda}^{\mathrm{v}''}(\tau, \bm{k})
  + \left( k^{2} - \frac{a''}{a} \right)
  h^{\mathrm{v}}_{\lambda}(\tau, \bm{k}) = 0,
\end{equation}
with the initial conditions given by the Bunch-Davies vacuum,
\begin{equation}
  \left. h^{\mathrm{v}}_{\pm}(\tau,\bm{k}) \right|_{- k \tau \gg 1}
  = \frac{1}{\sqrt{2k}} \MaE^{ - \MaI k \tau }.
\end{equation}
On the other hand, $\hat{h}_{\lambda}^{\mathrm{s}}(\tau, \bm{k})$ is a non-homogeneous solution with homogeneous boundary conditions, corresponding to the induced GWs. After performing the Fourier transform, one obtains
\begin{equation}
\label{eq: hq}
\begin{aligned}
  & h_{\lambda}^{\mathrm{s}''}(\tau, \bm{k})
  + \left( k^{2} - \frac{a''}{a} \right) {h}_{\lambda}^{\mathrm{s}}(\tau, \bm{k}) \\
  &= -\frac{a^{3}}{M_{\mathrm{pl}}} \left( \varepsilon_{i}^{\lambda}(\hat{k}) \varepsilon_{j}^{\lambda}(\hat{k}) \right)^{\!*}
  \int \frac{\mathrm{d}^{3} p}{(2\pi)^{3/2}} \\
  &\times \left( \hat{E}_{i}(\tau,\bm{p}) \hat{E}_{j}(\tau,\bm{k}-\bm{p})
      + \hat{B}_{i}(\tau,\bm{p}) \hat{B}_{j}(\tau,\bm{k}-\bm{p}) \right),
\end{aligned}
\end{equation}
which can be solved by the Green's function method \cite{Cook:2013xea, Sorbo:2011rz},
\begin{align}
  \notag
  \hat{h}_{\lambda}^{\mathrm{s}}(\tau, \bm{k})
  & = - \dfrac{1}{M_{\mathrm{pl}}}
  \int_{-\infty}^{\tau} \mathrm{d} \tau' \frac{G_{k}(\tau, \tau')}{a(\tau')}
  \int \dfrac{\mathrm{d}^{3} \bm{q}}{(2 \MaPI)^{3 / 2}}\\
  \notag
& \times \varepsilon_{l}^{-\lambda}(\hat{k}) \varepsilon_{m}^{-\lambda}(\hat{k})
  \Bigl(
    \tilde{A}'_{l}(\tau' ,\bm{q}) \hat{A}_{m}'(\tau' ,\bm{k} - \bm{q}) \\
  \label{eq: sourced_tensor}
    & - \varepsilon_{lab} q_{a} \hat{A}_{b}(\tau',\bm{q})
    \varepsilon_{mcd}( k_{c} - q_{c}) \hat{A}_{d} (\tau',\bm{k} - \bm{q})
  \Bigr),
\end{align}
where the Green's function is given by Eq.~\eqref{eq: green_func}.

The power spectrum of GWs, $\mathcal{P}^{\lambda}_{h}(\tau, k)$, is defined by
\begin{equation}
  \braket{\hat{h}_{\lambda}(\tau, \bm{k})\hat{h}_{\lambda}(\tau, \bm{k'})}
  \equiv \dfrac{2 \MaPI^{2}}{k^{3}}\dfrac{a(\tau)^2}{4}\mathcal{P}^{\lambda}_{h}(\tau,k) \delta(\bm{k} + \bm{k}').
\end{equation}
By applying Wick's theorem and retaining only the enhanced mode, the power spectrum is given by~\cite{Garcia-Bellido:2023ser}
\begin{align}
\notag
  & \mathcal{P}_{h}^{\lambda}(\tau,k)
  = \frac{H^{2}}{\MaPI^{2} M_{\mathrm{pl}}^{2}}
  + \frac{k^{3}}{\MaPI^{4} M_{\mathrm{pl}}^{4}} \int_{0}^{\infty} q^{2} \mathrm{d} q \\
\notag
  &\times \int_{-1}^{1} \mathrm{d} u
  \Biggr[
    \left| \epsilon^{\lambda}_{i}(\hat{k}) \epsilon^{-}_{i}(-\hat{q}) \right|^{2}
    \left| \epsilon^{\lambda}_{j}(\hat{k}) \epsilon^{-}_{j}(\frac{\bm{q}-\bm{k}}{|\bm{q}-\bm{k}|}) \right|^{2} \\
  &\times \biggl|
  \int_{-\infty}^{0} \mathrm{d} \tau' \frac{G_{k}(\tau, \tau')}{a(\tau)a(\tau')}
  \biggl(
    A^{'-}(\tau',\bm{q}) A^{'-}(\tau', \bm{k}-\bm{q})\notag\\
\label{eq: Ph}
  &+ q |\bm{k} - \bm{q}|
    A^{-}(\tau',\bm{q})
    A^{-}(\tau',\bm{k} -\bm{q}) \biggr) \biggr|^{2} \Biggr],
\end{align}
where $u \equiv \cos \theta$, and the norms of the polarization vector can be computed using
\begin{equation}
  \left| \epsilon_{i}^{\lambda}(\hat{p})
    \epsilon_{i}^{\lambda'}(\hat{q}) \right|^{2}
  = \left( \frac{1 - \lambda \lambda' \hat{p} \cdot \hat{q}}{2} \right)^{2}.
\end{equation}
Today’s energy spectrum of the stochastic GW background is related to the GW power spectrum at the end of inflation by \cite{Caprini:2018mtu}
\begin{equation}
\label{eq: omega_GW}
  \Omega_{\mathrm{GW},0}h^2 = \frac{\Omega_{\mathrm{r}, 0} h^{2}}{24}
  ( \mathcal{P}^{+}_{h}(\tau_{\mathrm{end}},k) + \mathcal{P}^{-}_{h}(\tau_{\mathrm{end}},k) ),
\end{equation}
where $\Omega_{\mathrm{r}, 0} \simeq 4.15 \times 10^{-5} / h^{2}$ denotes the current fraction of the radiation energy density to the total energy density of the Universe.

\section{Model}
\label{sec: implementation}

In this section, we present an implementation of the mechanism introduced in Sec.~\ref{sec: mechanism}. In our implementation, the potential of the inflaton is the Starobinsky potential,
\begin{equation}
  \label{eq:V_potential}
  V(\phi) = V_{0} \left( 1 - \MaE^{-\sqrt{\frac{2}{3}}\frac{\phi}{M_{\mathrm{pl}}} } \right)^2,
\end{equation}
and the coupling $U(\phi, \chi)$ has the form
\begin{equation}
  \label{eq: U_g}
  U(\phi,\chi)\text{=}\frac{1}{2}m_{\chi}^2 (\chi -g(\phi ))^2.
\end{equation}
The parameter $m_{\chi}$ here is consistent with our previous definition in Eq.~\eqref{eq: m_chi}, since the second-order partial derivative yields $U_{,\chi\chi}(\phi, \chi) = m_{\chi}^{2}$. Moreover, the order of magnitude of the coupling is much smaller than the inflaton potential, $U(\phi, \chi) \ll V(\phi)$, and therefore the dynamics of the inflaton are governed by $V(\phi)$. For this system, the background equations in Eqs.~\eqref{eq: phi_eom} and \eqref{eq: chi_eom} become
\begin{align}
  \label{eq: phi_eom_realize}
  & \ddot{\phi}_{0} + 3 H \dot{\phi}_{0} + V_{,\phi}(\phi_{0})
  - m_{\chi}^{2} (\chi - g(\phi_{0})) g_{,\phi}(\phi_{0}) = 0, \\
  \label{eq: chi_eom_realize}
  & \ddot{\chi}_{0} + 3 H \dot{\chi}_{0} + m^{2}_{\chi}(\chi_{0} - g(\phi_{0}))
  = \frac{\alpha}{f} \braket{\bm{E} \cdot \bm{B}}.
\end{align}
In the large mass limit, $m_{\chi} \gg H$, the differential equation in Eq.~\eqref{eq: chi_eom_realize} reduces to an algebraic equation, $\chi_{0} - g(\phi_{0}) = 0$, which implies that the evolution of the axion $\chi_{0}(t)$ is governed by the function $g(\phi)$ via $\chi_{0}(t) = g(\phi_{0}(t))$. Therefore, $\chi_{0}$ and $g(\phi_{0})$ have similar orders of magnitude. Physically, the function $g(\phi_{0})$ encodes how the inflaton $\phi$ “drags” the axion $\chi$ through the coupling $U(\phi, \chi)$. However, the deviation between $\chi_{0}(t)$ and $g(\phi_{0}(t))$ can become significant when $g(\phi_{0}(t))$ varies rapidly, as $\chi_{0}(t)$ must move significantly to catch up with $g(\phi_{0}(t))$. As a result, $\chi_{0}(t)$ experiences rapid motion near regions in field space where $g_{,\phi}(\phi_{0})$ is large. In these moments, the rapidly rolling axion triggers the tachyonic instability of the gauge field mode functions, leading to strong gauge particle production as described in Sec.~\ref{sec: dynamics}.

\begin{figure}[tbp]
  \includegraphics[width=.5\textwidth]{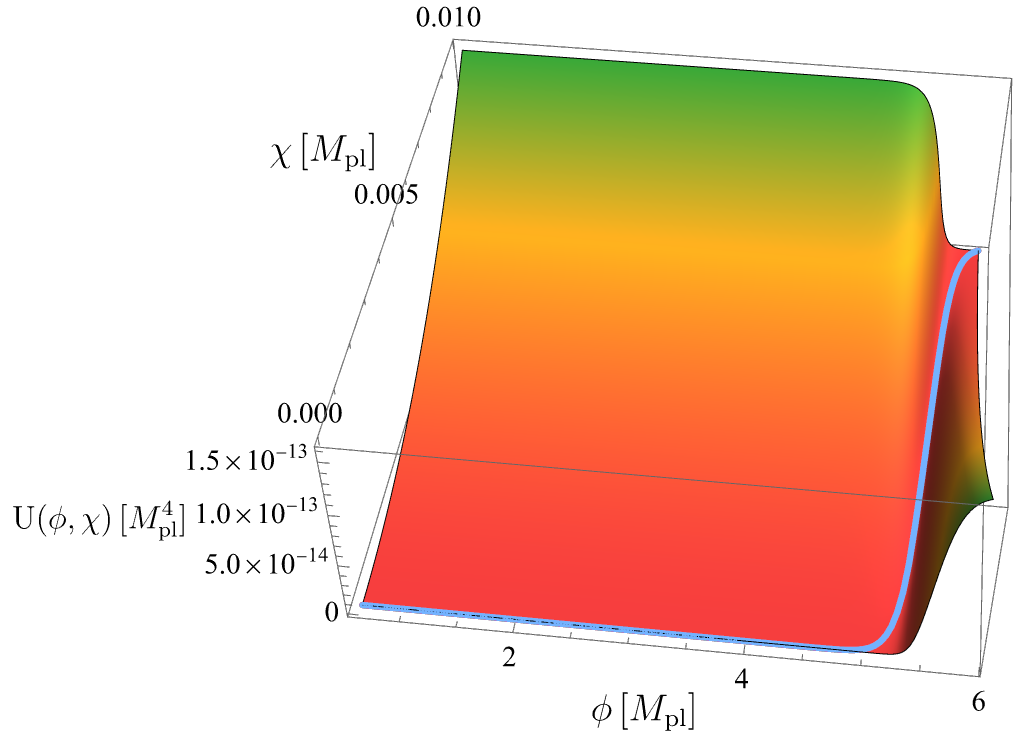}
  \caption{
  Landscape of the coupling~\eqref{eq: U_g}. The light blue trajectory represents the solution of Eq.~\eqref{eq: phi_eom_realize} and Eq.~\eqref{eq: chi_eom_realize}. This trajectory includes a rapid turn as it nearly reaches the $\phi$-axis, at which point gauge quanta are exponentially produced due to the tachyonic instability.
  }
  \label{fig: trajectory_potential}
\end{figure}

In our study, we choose the parameters as $V_0 = 9.95 \times 10^{-11} M_{\mathrm{pl}}^4$, $m_{\chi} = 5.8 \times 10^{-5} M_{\mathrm{pl}}$, the decay parameter is set to $f = 7.16 \times 10^{-5} M_{\mathrm{pl}}$, and the coupling constant is taken as $\alpha = 1$, resulting in $U(\phi,\chi)\ll V(\phi)$ throughout the evolution.
These choices ensure that $m_{\chi}^{2} \sim \mathcal{O}(100 \, H^{2})$, meaning the perturbation $\delta \chi$ is never excited. As a result, the conversion term $J_{\mathrm{D}} Q_{\chi}$ in Eq.~\eqref{eq: Qphi_perturbation} is negligible compared to the Hubble term $H^{2} Q_{\phi}$. For the function $g(\phi)$, which controls the motion of the axion, we only require that it induces a period of rapid rolling. In our numeric computation, we choose the following function
\begin{equation}
  g(\phi) = \Lambda_{0} \arctan\left[ \exp \left(
    \delta\, (\phi-\phi_{*})
  \right)\right]
  \label{eq: g_phi}
\end{equation}
with $\delta = 10 \, M_{\mathrm{pl}}^{-1}$, $\Lambda_{0} = 6.32 \times 10^{-3} \, M_{\mathrm{pl}}$, and $\phi_{} = 5.48 \, M_{\mathrm{pl}}$. 
 The parameter $\phi_{*}$ controls the time when the axion starts fast rolling, which corresponds to the CMB scale in this study. Using the function in Eq.~\eqref{eq: g_phi}, the motion of the axion $\chi(t) = g(\phi(t))$ resembles the slow-roll solution of an axion with a cosine potential~\cite{Namba:2015gja,Khlopov:1998uj}. Our choice of $\Lambda_{0}$ ensures that $m_{\phi}^{2} = m_{\chi}^2 g_{,\phi}(\phi_{0})^2 \ll H^2$, making the influence of the effective potential $U(\phi, \chi)$ on the background evolution $\phi_{0}(t)$ negligible. The landscape of the coupling $U(\phi, \chi)$ is illustrated in Fig.~\ref{fig: trajectory_potential}.

\begin{figure}[tbp]
  \includegraphics[width=.5\textwidth]{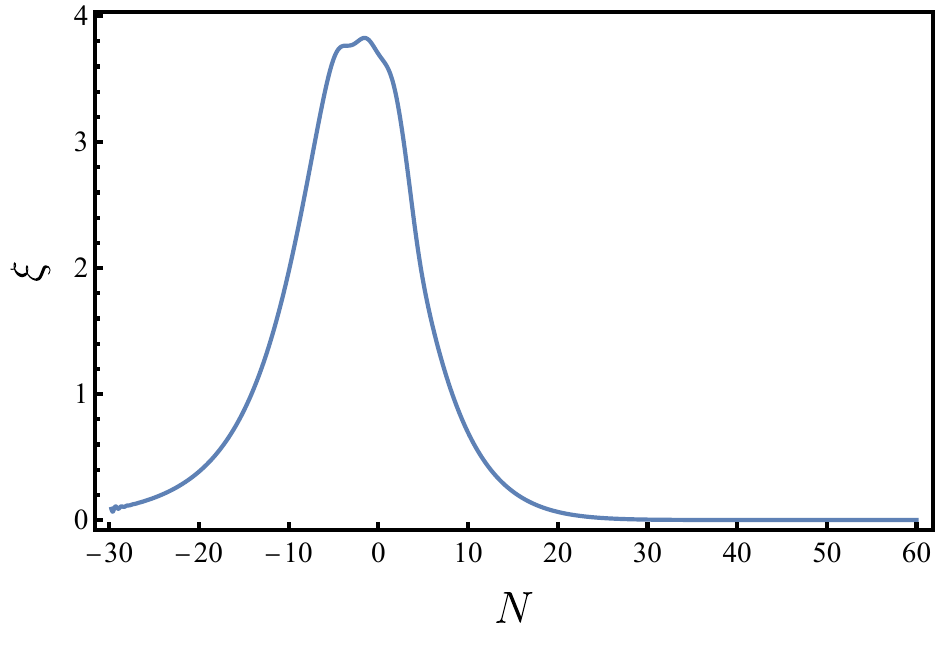}
  \caption{
  Evolution of $\xi$ as a function of $N$. In the region where $\xi \gg 1$, the axion field $\chi$ enters a fast-roll phase, leading to the exponential production of gauge quanta.
  }
  \label{fig: xi}
\end{figure}

\begin{figure}[tbp]
  \includegraphics[width=.5\textwidth]{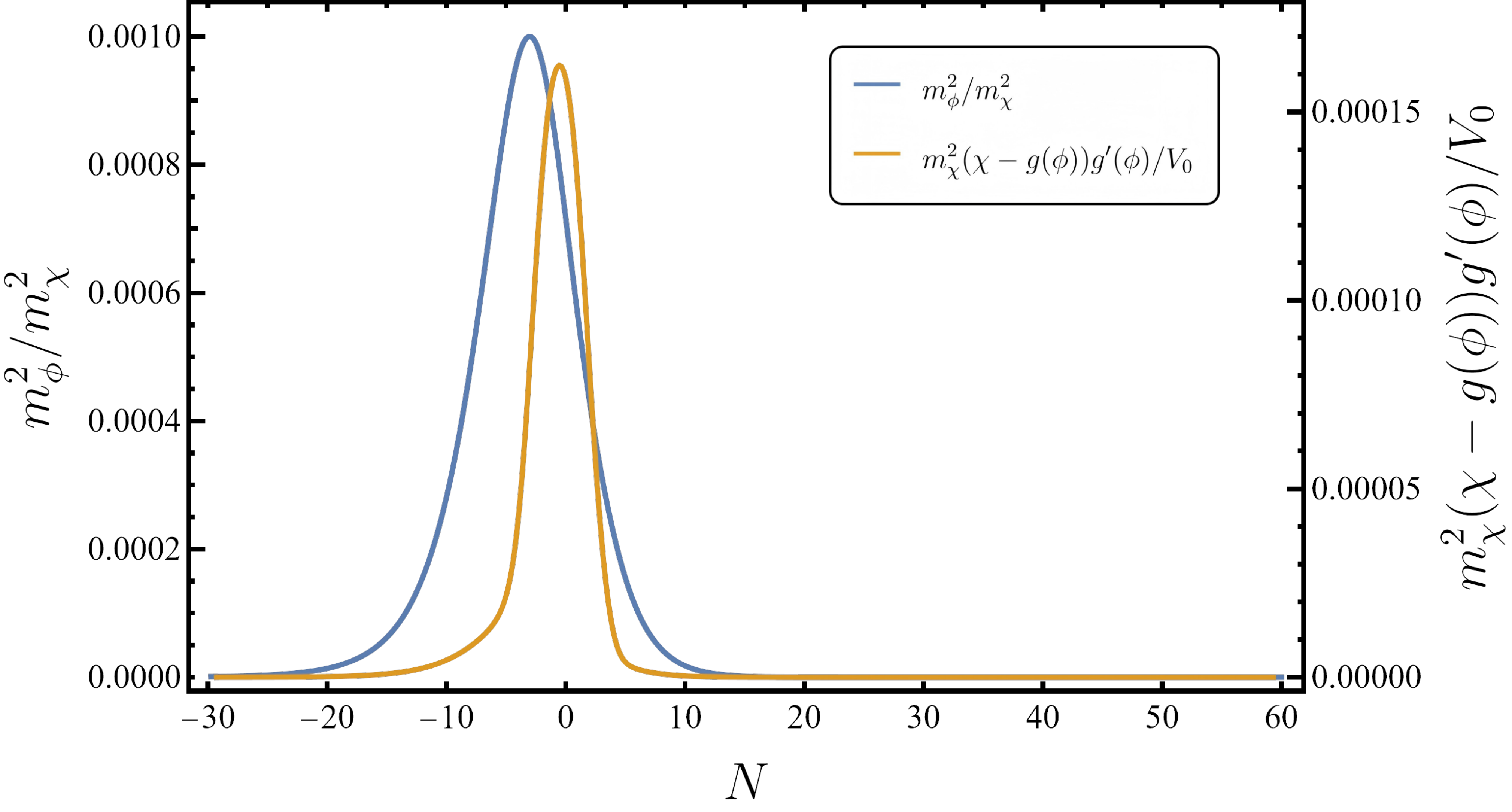}
  \caption{
  Evolution of the coupling term $m_{\chi}^{2} (\chi - g(\phi)) g_{,\phi}(\phi)$ in Eq.~\eqref{eq: phi_eom_realize} and the mass ratio $m_{\phi}^2/m_{\chi}^2$.
  }
  \label{fig: phi_friction_due_chi}
\end{figure}

Our numerical computation begins with the initial values $\phi_{\mathrm{ini}} = 5.9 \, M_{\mathrm{pl}}$ and $\chi_{\mathrm{ini}} = g(\phi_{\mathrm{ini}})$. The initial value of $\chi_0$ ensures that it rolls from a partial minimum, where $U_{,\chi}(\phi_{0}, \chi_{0}) = 0$. The end of inflation is determined by the condition $\epsilon_{H} \equiv -\dot{H}/{H^{2}} = 1$. Additionally, we set the e-folding number $N = 0$ at the time when the pivot scale $k_{*} = 0.05 \, \mathrm{Mpc}^{-1}$ exits the horizon, and $N = 60$ at the end of inflation.
The numerical method used in this paper follows our previous works \cite{He:2024bno, He:2025ieo, Zhang:2025cyd}, where we evolve both the background quantities in Eqs.~\eqref{eq: a_eom_1}-\eqref{eq: chi_eom} and the gauge field in Eq.~\eqref{eq: Aq} simultaneously. At each iteration, we compute the backreaction terms by integrating over the gauge field in Fourier space using Eqs.~\eqref{eq: ensemble_EB}-\eqref{eq: ensemble_BB}.
The trajectory of the solution in the $U(\phi, \chi)$ landscape is shown as a blue curve in Fig.~\ref{fig: trajectory_potential}, where a rapid turn occurs as the field rolls down and nearly reaches the $\phi$-axis. This rapid turn leads to a brief fast-roll stage of the axion, triggering the tachyonic instability of the gauge field. We plot the evolution of the parameter $\xi$ in Fig.~\ref{fig: xi}, where the bump indicates the time when the fast-roll occurs. Due to the small value of $\xi$, the backreaction has no significant influence on the system. In Fig.~\ref{fig: phi_friction_due_chi}, we show the evolution of the mass ratio $m_{\phi}^{2} / m_{\chi}^{2}$ and the gradient term $|U_{,\phi}| = m_{\chi}^{2}(\chi - g(\phi)) g_{\chi}(\phi_{0})$. This figure demonstrates that, although these terms increase at the turning point, they remain small enough to have no significant effect on the evolution of the inflaton.

\begin{figure}[tbp]
  \includegraphics[width=.5\textwidth]{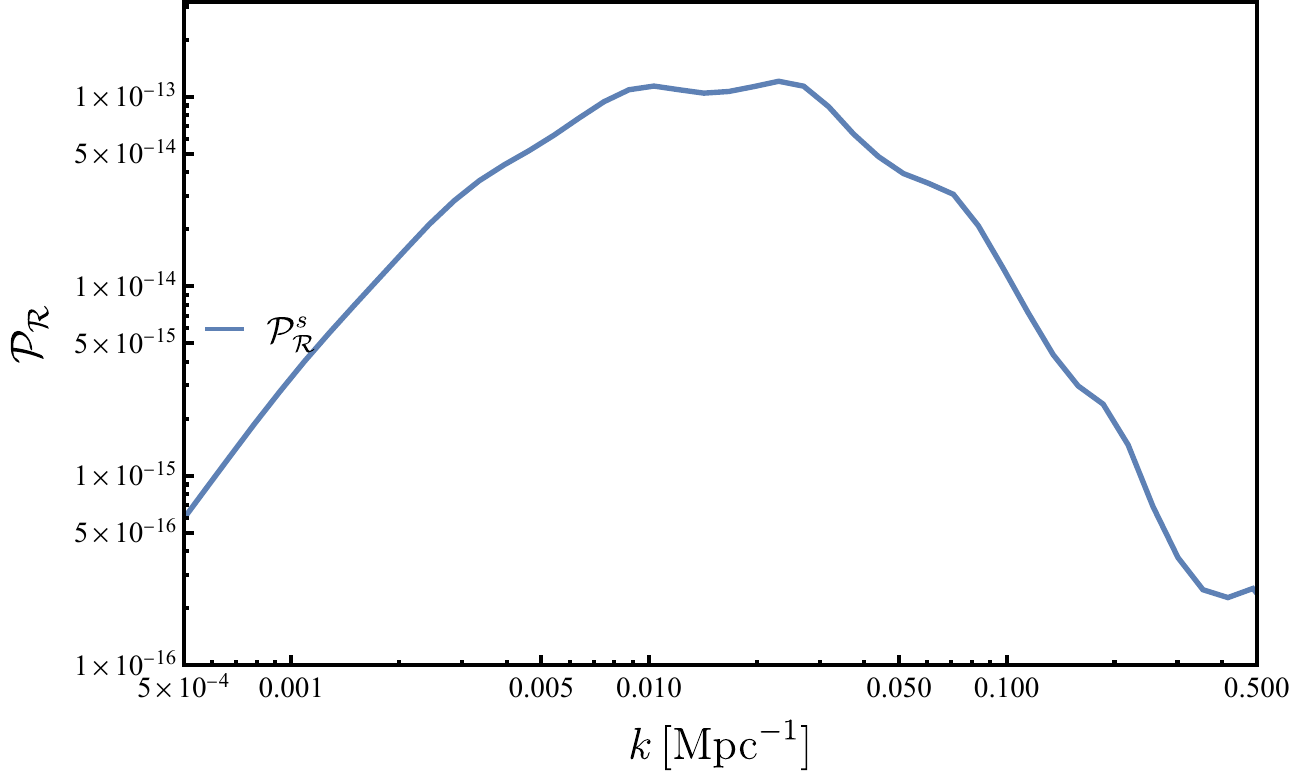}
  \caption{
  Sourced scalar power spectrum $\mathcal{P}_{\mathcal{R}}^{\mathrm{s}}(\tau_{\mathrm{end}}, k)$ on CMB scales. It is computed numerically using Eq.~\eqref{eq: PR}. The induced peak amplitude is approximately $10^{-4}$ times smaller than the amplitude of vacuum power spectrum, $\mathcal{P}_{\mathcal{R}}^{\mathrm{v}} \approx 2.1 \times 10^{-9}$, on CMB scales.
  }
  \label{fig: CMB_scalar}
\end{figure}

\begin{figure}[tbp]
  \includegraphics[width=.5\textwidth]{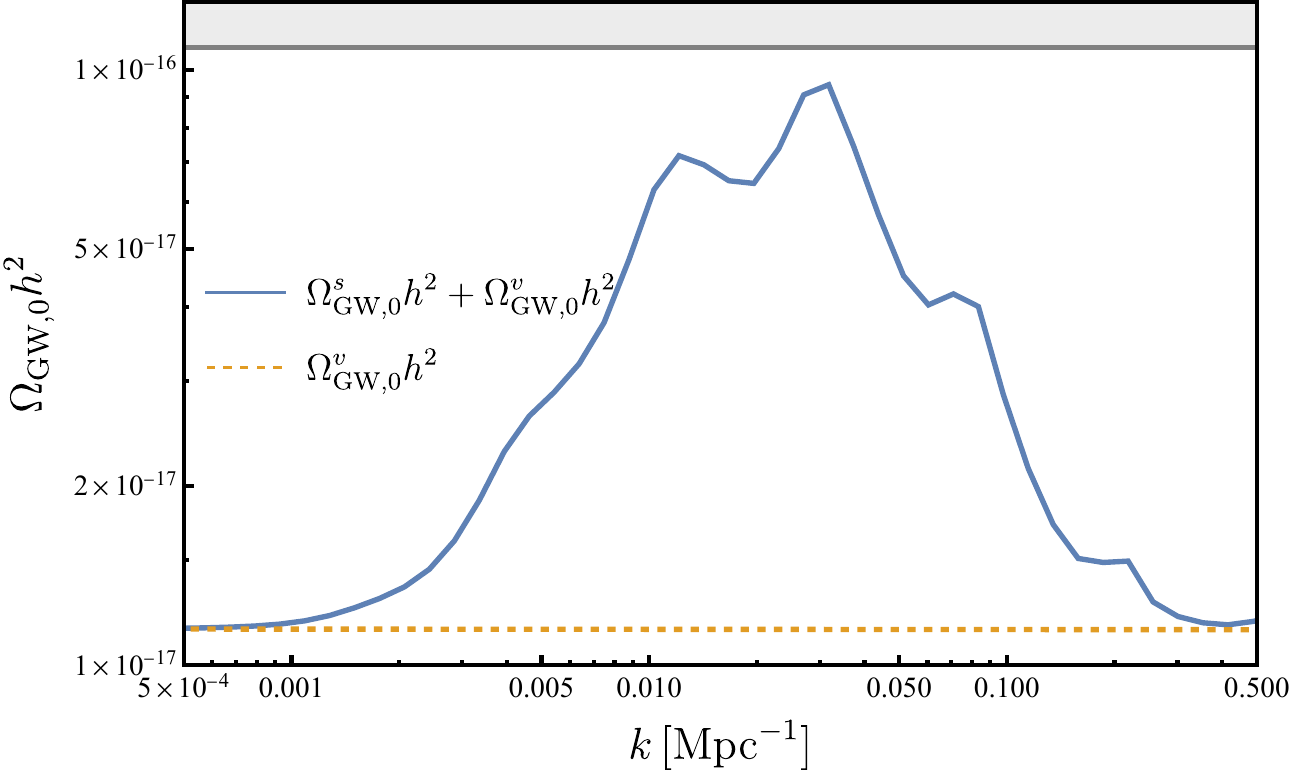}
  \caption{
  Energy spectrum of GWs on CMB scales, computed from Eq.~\eqref{eq: Ph} and Eq.~\eqref{eq: omega_GW}. The total GW signal, denoted by the blue curve, is potentially detectable for a large tensor-to-scalar ratio $r \approx 0.026$, which is close to the upper bounds from BICEP3+ and Planck PR4+~\cite{Tristram:2021tvh, Galloni:2022mok}, shown in gray. For comparison, the vacuum contribution is also plotted as a dashed orange line.
 }
  \label{fig: CMB_gw}
\end{figure}

We then compute the spectrum of both scalar and tensor perturbations sourced by gauge quanta based on Eq.~\eqref{eq: PR} and Eq.~\eqref{eq: omega_GW}. Fig.~\ref{fig: CMB_scalar} shows the power spectrum of curvature perturbations sourced by the excited gauge field. Notably, its maximum value is significantly smaller than the vacuum contribution in this implementation, with $\mathcal{P}_{\mathcal{R}}^{\mathrm{v}} \sim 10^{-9}$. Fig.~\ref{fig: CMB_gw} displays the total energy density of the GWs. An important feature is that the energy density of the sourced GWs is approximately one order of magnitude larger than the vacuum contribution, placing it well within the sensitivity range of the CMB-S4 project ($r > 0.003$ at greater than $5\sigma$)~\cite{CMB-S4:2020lpa}.

\section{Conclusions}
\label{sec: conclusion}
In this paper, we propose a mechanism that suppresses scalar perturbations in axion-gauge field dynamics.
Our approach employs a multi-field framework consisting of a heavy pseudo-scalar field initialized in its vacuum, a slowly rolling inflaton, and a U(1) gauge field. The heavy field is dragged by the inflaton along the minima of the coupling $U(\phi, \chi)$. When the trajectory in field space undergoes a sharp turn, the heavy field enters a brief fast-roll phase, which induces a tachyonic instability through the Chern-Simons interaction. The amplified gauge field in turn sources both GWs and curvature perturbations.
Compared to the conventional axion spectator model~\cite{Namba:2015gja, Campeti:2022acx, Ozsoy:2017blg, Mukohyama:2014gba, Barnaby:2012xt}, our approach significantly suppresses scalar perturbations by removing the dominant mixing channel $\delta \chi \to \delta \phi$.
As a concrete demonstration, we show that our mechanism can generate a potentially observable GW signal on CMB scales without violating CMB observational constraints - a result unattainable in standard inflationary models containing an axion field~\cite{Barnaby:2010vf, Barnaby:2011vw, Barnaby:2011qe, Linde:2012bt, Planck:2015zfm, Ozsoy:2014sba}.
Typically, constraints from scalar perturbations prevent the gauge fields from being sufficiently amplified to produce a detectable GW signal.
However, in our framework, the mode function of the heavy field remains unexcited due to its large mass. Consequently, the dominant source of curvature perturbations - the direct Chern-Simons interaction - is replaced by purely gravitational effects. This leads to scalar perturbations that are significantly suppressed relative to the vacuum contribution (see Fig.~\ref{fig: CMB_scalar}).
Meanwhile, tensor perturbations remain unaffected, allowing for a substantially enhanced tensor-to-scalar ratio.
This enhancement places our model within the discovery potential of upcoming CMB experiments (see Fig.~\ref{fig: CMB_gw}), providing a promising and testable avenue for probing axion-gauge field dynamics in the early Universe.

In principle, our mechanism can be applied to a wide range of particle-production scenarios~\cite{Senatore:2011sp, Barnaby:2012xt}.
The essential ingredient is a heavy field as the particle-production sector, with its dynamics controlled by an effective potential that depends on the light field.
With appropriate parameter choices, the dynamics of the light field can remain unaffected at both the background and perturbation levels.
Because the heavy field remains massive, scalar perturbations from its direct couplings to particles are suppressed, leaving only gravitational couplings. It is emphasized that the multi-field potential with a curved trajectory of minima need not take the specific form of Eq.~\eqref{eq: U_g}. The only requirement is that the heavy field remains massive throughout inflation.
In fact, different choices of the potential can satisfy different requirements.

\begin{acknowledgments}
This work is supported in part by the National Natural
Science Foundation of China under Grants No. 12475067,
No. 12305057, and No. 12235019.
\end{acknowledgments}

\bibliography{reference} 

\appendix

\end{document}